\documentclass[10pt,journal]{IEEEtran}

\usepackage[english]{babel}
\usepackage[usenames]{color}
\usepackage{colortbl}
\usepackage{comment}
\usepackage{graphicx}
\usepackage{epsfig}
\usepackage{array, colortbl}
\usepackage{listings}
\usepackage{epstopdf}
\usepackage{multirow}
\usepackage{rotating}
\usepackage{float}
\usepackage[obeyspaces,hyphens,spaces]{url}
\usepackage{balance}
\usepackage{fancybox}
\usepackage{scalefnt}
\usepackage[normalem]{ulem}
\usepackage{cleveref}
\usepackage{xspace}
\usepackage{subcaption}
\usepackage{dingbat}
\usepackage{booktabs}

\usepackage[numbers]{natbib}
\makeatletter
\def\NAT@spacechar{~}%
\makeatother

\usepackage{nameref}
\usepackage{soul}
\usepackage[dvipsnames]{xcolor}

\ifCLASSINFOpdf
\else
\fi

\usepackage{enumitem}
\setlist[enumerate]{leftmargin=*}

\definecolor{amber}{rgb}{1.0, 0.49, 0.0}
\definecolor{lavenderindigo}{rgb}{0.58, 0.34, 0.92}
\definecolor{islamicgreen}{rgb}{0.0, 0.56, 0.0}

\newcommand{\npm}{\texttt{npm}\xspace}
\newcommand{\npms}{\textit{npms}\xspace}
\newcommand{\node}{\textit{Node.js}\xspace}

\newcommand{\stable}{not in decline\xspace}
\newcommand{\decline}{in decline\xspace}

\newcommand*\fullcaption[2][]{\caption[#1]{#1\xspace#2}}

\usepackage[colorinlistoftodos,disable]{todonotes}
\usepackage{verbatim}

\makeatletter
\if@todonotes@disabled
\else
\paperwidth=\dimexpr \paperwidth + 6cm\relax
\oddsidemargin=\dimexpr\oddsidemargin + 3cm\relax
\evensidemargin=\dimexpr\evensidemargin + 3.5cm\relax
\marginparwidth=\dimexpr \marginparwidth + 3.5cm\relax
\fi
\makeatother

\makeatletter
\newcommand{\labelname}[1]{%
  \def\@currentlabelname{#1}}%
\makeatother

\newcommand\mainpoint[1]{\vspace{.1in}\noindent\textbf{#1.}}
\newcommand\saybox[1]{\begin{center}\noindent\textit{``#1''}\end{center}}

\usepackage[skins]{tcolorbox}
\newcommand{\conclusion}[1]{\begin{center}\begin{tcolorbox}[skin=widget, left=2mm,right=2mm,top=2mm,bottom=2mm,boxrule=0.3mm,arc=0mm,coltitle=black,colframe=black!99!white,colback=white!88!gray,width=(\linewidth),before=\hfill,after=\hfill]#1\end{tcolorbox}\end{center}}

\usepackage{pifont}

\usepackage{tikz}
\newcommand*\circled[1]{\tikz[baseline=(char.base)]{\node[shape=circle,draw,inner sep=.5pt] (char) {#1};}}

\makeatletter
\renewcommand{\paragraph}[1]{\noindent\textsf{#1}.}

\begin{document}

\title{Towards Using Package Centrality Trend to Identify Packages in Decline}%

\author{Suhaib~Mujahid,
  Diego~Elias~Costa,
  Rabe~Abdalkareem,
  Emad~Shihab, %
  Mohamed~Aymen~Saied, and
  Bram~Adams%

  \IEEEcompsocitemizethanks{
    \IEEEcompsocthanksitem
    S. Mujahid, D. E. Costa, and E. Shihab
    are with the Data-driven Analysis of Software (DAS) Lab
    at the Department of Computer Science and Software Engineering,
    Concordia University,
    Montreal, Canada.
    Email:~\{suhaib.mujahid,~diego.costa,~emad.shihab\}@concordia.ca
    \IEEEcompsocthanksitem
    R. Abdalkareen
    is with the School of Computer Science,
    Carleton University,
    Ottawa, Canada.
    Email: rabe.abdalkareem@carleton.ca
    \IEEEcompsocthanksitem
    M.A. Saied
    is with the Department of Computer Science and Software Engineering,
    Laval University,
    Quebec City, Canada.
    Email:~\mbox{mohamed-aymen.saied@ift.ulaval.ca}
    \IEEEcompsocthanksitem
    B. Adams
    is with the School of Computing,
    Queen's University,
    Kingston, Canada.
    Email: bram.adams@queensu.ca
  }

  \thanks{Manuscript accepted October 18, 2021.}

}

\markboth{IEEE Transactions on Engineering Management, October~2021}%
{Shell \MakeLowercase{\textit{et al.}}: Bare Advanced Demo of IEEEtran.cls for IEEE Computer Society Journals}

\IEEEtitleabstractindextext{
  \begin{abstract}

    Due to their increasing complexity, today's software systems are frequently built by leveraging reusable code in the form of libraries and packages.
    Software ecosystems (e.g., \npm) are the primary enablers of this code reuse, providing developers with a platform to share their own and use others' code.
    These ecosystems evolve rapidly: developers add new packages every day to solve new problems or provide alternative solutions, causing obsolete packages to decline in their importance to the community.
    Developers should avoid depending on packages in decline, as these packages are reused less over time and may become less frequently maintained.
    However, current popularity metrics (e.g., Stars, and Downloads) are not fit to provide this information to developers because their semantics do not aptly capture shifts in the community interest.

    In this paper, we propose a scalable approach that uses the package's centrality in the ecosystem to identify packages in decline.
    We evaluate our approach with the \npm ecosystem and show that the trends of centrality over time can correctly distinguish packages in decline with an ROC-AUC of 0.9.
    The approach can capture 87\% of the packages in decline, on average 18 months before the trend is shown in currently used package popularity metrics.
    We implement this approach in a tool that can be used to augment the \npms metrics and help developers avoid packages in decline when reusing packages from \npm.

  \end{abstract}

  }

\maketitle

\IEEEdisplaynontitleabstractindextext

\IEEEpeerreviewmaketitle

\section{Introduction}
\label{sec:centrality:introduction}
\IEEEPARstart{S}{oftware} ecosystems have changed the way we develop software.
Platforms like \npm, PyPI, and Maven enable developers to easily reuse code from other projects in the form of packages~\cite{Kikas_MSR2017}, boosting development productivity~\cite{Abdalkareem_FSE2017}, and improving software quality~\cite{Zerouali_SANER2017}. As such, these ecosystems are becoming extremely popular and large. For example, the node package manager (\npm) alone has more than 1.4 million packages to date and is seeing exponential growth~\cite{Decan_EMSE2019}.

The large size and rapid evolution of these ecosystems has its downsides as well. For example, new (and often better) packages are continuously being introduced~\cite{Kula_EMSE2017, Wittern_MSR2016, Abdalkareem_EMSE2020,Besten_TEM2020}, making other, once popular packages, obsolete, dormant or even deprecated~\cite{Valiev_FSE2018}. As such, it is becoming increasingly important for application developers to ensure that they choose the right packages from the ecosystem.

Although prior work examined projects that are unmaintained~\cite{Coelho_ESEM2018, Coelho_IST2020}, to the best of our knowledge, little attention has focused on identifying packages that lose popularity over time (i.e., are \decline). At the same time, current popularity metrics that are commonly used by developers to select packages, such as downloads and stars, are not adequate to capture a shift in community interest.
For example, the number of downloads represents not only the number of times a package is installed on its own, but also the number of times it is installed as a dependency of other packages. Hence, the popularity of a dependent package could heavily impact the number of downloads of its dependencies~\cite{Dey_PROMISE2018}.
Also, the number of stars a package is linked to its repository, which may include many other packages and is unlikely to decrease to reflect interest shift over time~\cite{Zhou_FSE2019, Borges_JSS2018}.

Therefore, in this paper we use the package's centrality as a proxy of community interest. Community interest drives packages to evolve, i.e., include better features driven by community needs, keep up the package maintenance by reporting bugs to maintainers, motivate maintainers to continue supporting the package, and some times even financially support the maintainers on platforms such GitHub Sponsors,\footnote{https://github.com/sponsors} Open Collective,\footnote{https://opencollective.com} and Tidelift.\footnote{https://tidelift.com}
On the other hand, packages that are declining in community interest are reused less over time, may become less actively maintained, and in more extreme cases, even become abandoned~\cite{Valiev_FSE2018, Khondhu_OOS2013}.
Furthermore, the decline in community interest of a package may indicate that a better solution is drawing attention in the ecosystem, and developers are migrating to a package that better suits their needs.

Hence, our aim is to effectively identify packages that may be \decline. To do so, we use the package centrality to identify declining community interest.
By definition, centrality is a measure of the prominence or importance of a node in a social network~\cite{Wasserman_Book1994}.
Centrality has been used in many fields e.g.,
in finance to measure the stability of banks in financial networks~\cite{Wang_WI2017},
in electrical engineering to rank the importance of components in network infrastructures~\cite{Cadini,Stergiopoulos}, and other fields including computer science and software engineering~\cite{Maharani,Hong}.
In our context, centrality allows us to rank the packages based on the popularity/importance of packages that depend on them.
Specifically, we use the PageRank algorithm to evaluate the shift in their centrality over time.
The intuition is that packages that have a consistent decrease in the centrality ranking are likely to be packages \decline. Hence, package developers should be careful when depending on such packages.

We evaluate our approach on the \npm ecosystem. We do so since JavaScript is one of the most popular programming languages~\cite{StackOverflow_Survey:online} and \npm is the largest growing ecosystem to date~\cite{Decan_EMSE2019}. The popularity and scale of the \npm ecosystem makes it an ideal candidate for our study. We evaluate the effectiveness of using package centrality in identifying \npm packages that are in \decline. We formalize our study through the following research questions:
\begin{itemize}
	\item \nameref{rq:one}
	\item \nameref{rq:two}
	\item \nameref{rq:three}
\end{itemize}

Our findings show that our approach can detect 87\% of packages \decline with high accuracy, on average 18 months before current popularity metrics show the decline. Also, we find that our approach can detect packages \decline more than 16 months (on average) before such packages are deprecated. Lastly, we find that our approach complements commonly used popularity metrics such as dependents, downloads, stars, and forks when detecting packages \decline.

~\\
\noindent Our work makes the following contributions:
\begin{itemize}
	\item Propose a scalable approach to detect packages \decline using package centrality.
	\item Empirically evaluate our approach on the \npm ecosystem.
	\item Create a tool prototype to facilitate the usage of our approach by practitioners.
	\item Make all of our dataset (i.e., the collected data, analysis results, scripts) publicly available to support replication and future research~\cite{datasetonline}.
\end{itemize}

\mainpoint{Paper organization}
The remainder of this paper is organized as follows:
We motivate our work in \Cref{sec:centrality:motivation} with an example of a popular package \decline.
\Cref{sec:centrality:approach} details our approach, from data collection to computing centrality trends to find packages \decline.
In \Cref{sec:centrality:dataset} we explain how we collect and curate the baselines we use to evaluate our approach.
\Cref{sec:centrality:results} presents the findings of our empirical study by answering our three research questions.
We present a tool prototype to utilize our approach in \Cref{sec:centrality:implications}.
\Cref{sec:centrality:related_work} discusses the related work and
\Cref{sec:centrality:threats_to_validity} describes the threats to validity.
Finally, we conclude in \Cref{sec:centrality:conclusion}.

\section{Motivating Example}
\label{sec:centrality:motivation}

\newcommand{\moment}{Moment.js\xspace}

To illustrate the idea of using package centrality in determining a shift in community interest, we present the example of the \moment package.
\moment is a JavaScript library for parsing, validating, manipulating, and formatting dates.
This is a highly-used package, used in more than 1 million websites, including major companies\footnote{Reported by wappalyzer.com in January 2021} such as CNN, Microsoft Teams, LinkedIn and Dropbox.
\moment was developed using a now old-fashioned JavaScript packaging method, including all its functionalities in a single bloated JavaScript class.
Consequently, all websites that use \moment have to include the entire package regardless of the feature used, which incurs in an unnecessary overhead for website applications~\cite{moment_project_status}.

\saybox{Moment was built for the previous era of the JavaScript ecosystem. The modern web looks much different these days.}

Since 2018, alternatives to \moment (e.g.~date-fns and Day.js) have become more and more popular by providing similar functionality without incurring the overhead that \moment incurs.
Hence, the \npm community started shifting towards using more lightweight packages.
This shift includes migrating well-established open source projects like Google Chrome's Lighthouse, Vault by HashiCorp, and Web Stories by Google from \moment to other alternative packages~\cite{replace_moment:GoogleChrome,replace_moment:HashiCorp,replace_moment:google}.

\begin{figure}
    \centering
    \includegraphics*[trim=0 0 0 .20in, clip, width=.85\linewidth]{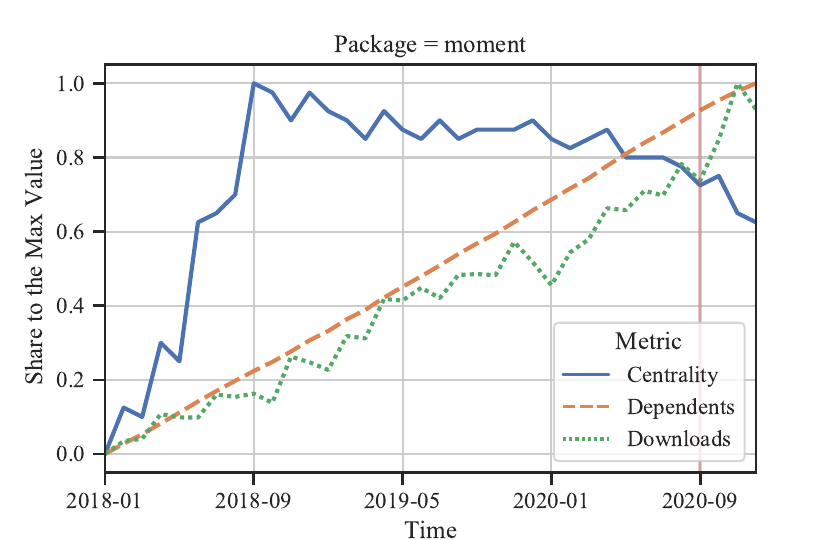}
    \fullcaption[Evolution of the \moment package on the centrality (PageRank), the number of downloads, and the number of dependents.]{
        The red vertical line indicates the time where maintainers reported \moment is now a legacy project.
        We normalize metric values using the min-max method where values range from 0 and 1~\cite{Codecademy_Normalization:online}.
    }
    \label{fig:motivation_example_moment}
\end{figure}

The popularity of the alternative packages led to a consistent decrease in \moment's centrality in the ecosystem starting in September 2018, which can be seen clearly in \Cref{fig:motivation_example_moment}.
On September 15th, 2020, the maintainers of \moment issued a statement in the README file indicating that the package is now a legacy project.
While maintainers have committed to still maintain the project, they recommend that users choose a different package.

The community's shift from using one of \npm's most used packages to other alternatives was public knowledge; however, none of the common popularity metrics, including the metrics used by the \npm search engine (\npms) were able to capture this phenomenon.
In fact, the number of downloads of \moment continued to increase (as shown in \Cref{fig:motivation_example_moment}) as well as the number dependent packages. As of January 2021, the \npm registry shows that 49,544 \npm packages depend on \moment and it is downloaded more than 16 million times a week.
The only metric that showed \moment's important decrease in \npm was centrality, which started to decrease as early as October 2018, the same year that alternative packages started to become more popular.

There are a couple of possible reasons why the number of downloads and dependents did not capture the decline of \moment.
First, since thousands of projects already use \moment, it will continue to be downloaded every time any of these projects get installed. Even when these projects migrate to use alternative packages, it will take much longer to reflect on the number of downloads due to technical lag where developers take a long time to update their dependencies~\cite{Decan_ICSME2018}.
Second, as \npm continues its exponential growth, newly created packages may still depend on \moment and substitute the core community that has migrated to the alternative solutions.
Package centrality, calculated with PageRank, accounts for not only the sheer number of dependents, but the importance of dependents in the network, which aptly captures the decline of \moment.
This example motivated us to investigate if package centrality trends can be used to identify packages that have declined in the community interest.

\section{Approach}
\label{sec:centrality:approach}
In this section, we explain our approach that uses the trend in the package's centrality in the \npm ecosystem and detect packages \decline.

\begin{figure*}[t]
    \centering
    \includegraphics[trim=20cm 11.1cm 20cm 11.1cm, clip,width=.9\linewidth]{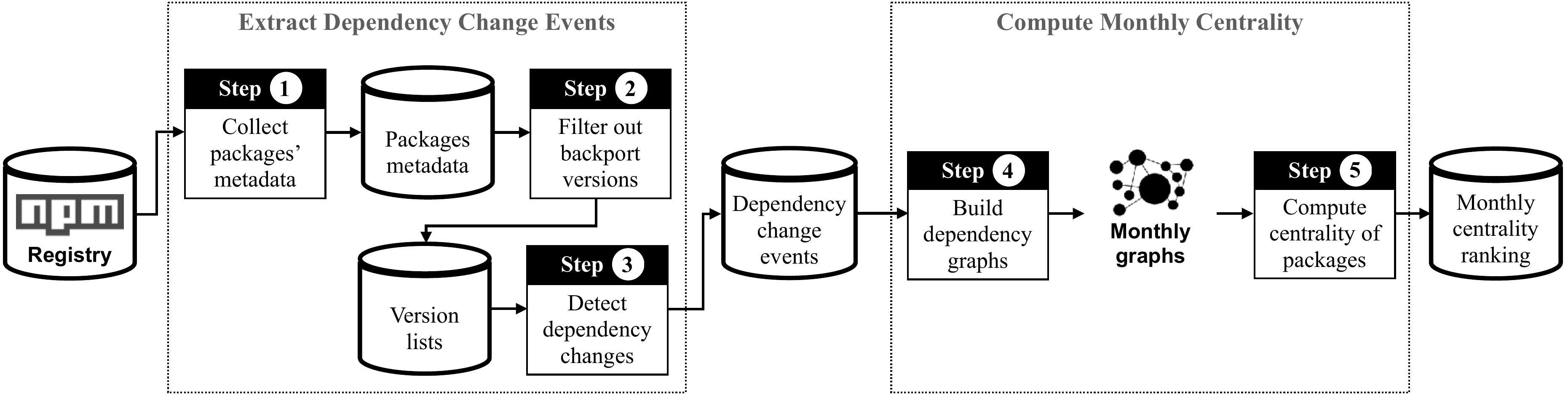}
    \caption{The approach to calculate centrality trends.}
    \label{fig:centrality_approach}
\end{figure*}

\subsection{Calculating Centrality Trends}
\label{sub:calculat_centrality}

Since the core idea of our approach is to use centrality, we need to efficiently calculate the centrality trends of packages. We first build a dependency graph containing all packages in \npm as nodes, and their dependency relationships as edges. We update this graph monthly with newly established dependencies and packages and compute the centrality metric for all packages. Each month, we rank the packages based on the value of their centrality metric.
In the following, we explain the attributes of our dependency graph, then, we describe our approach, illustrated in \Cref{fig:centrality_approach}, which includes how we:
i) collect and format the required metadata to build the dependency graph incrementally, and
ii) build the dependency graph each month to compute the centrality metric for all packages in the \npm ecosystem.

\begin{figure}
    \centering
    \includegraphics[trim=9.5cm 6.5cm 8cm 5cm, clip, width=\linewidth]{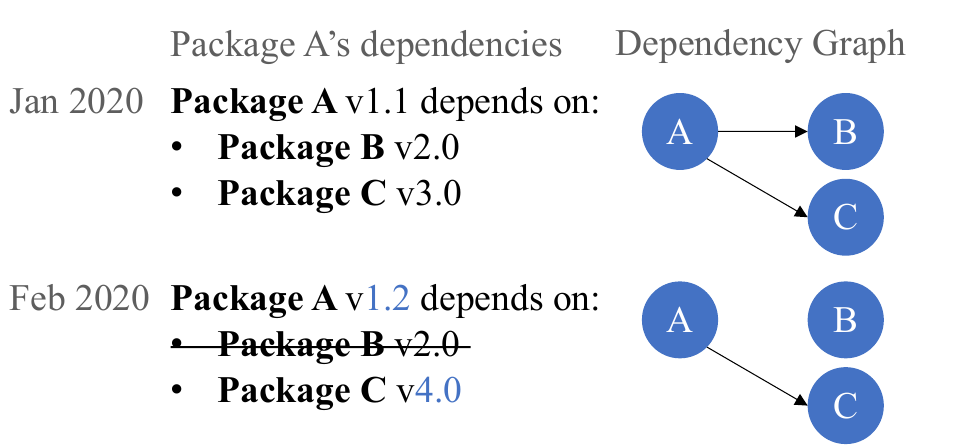}
    \caption{Illustration of our dependency graph build process}
    \label{fig:dependencygraph}
\end{figure}

\mainpoint{Attributes of Our Dependency Graph}
In order to use the package centrality as an indicator for packages in decline, our dependency graph needs to have two important properties:

\begin{enumerate}
    \item \textbf{Version insensitive nodes:}
          the nodes in our graph represent \npm packages, regardless of their versions.
          For instance, the popular package React has 298 distinct versions released in the \npm registry, but we represent it by only one node in our dependency graph.
          We do this because we are interested in capturing the usage shift without being affected by the technical lag in the dependency network, caused by developers taking a long time to update a dependency version~\cite{Decan_ICSME2018,Zerouali_ICSR2018}. %

    \item \textbf{Release sensitive edges:}
          an edge A~$\to$~B in our graph represents the dependency between the latest released version of package A on any version of package B.
          Once a new release of package A no longer depends on B, our dependency graph needs to reflect that by removing the A~$\to$~B edge. However, we do not consider backport versions as the latest released versions since they are not consistent with the package evolution time series.
\end{enumerate}

To better illustrate how this dependency graph is built, \Cref{fig:dependencygraph} presents an example of one package's dependencies and how they are reflected in our dependency graph.
As shown in \Cref{fig:dependencygraph}, the graph in each month (January and February) uses the latest version of Package A to add the edges from node A to its dependencies, but disregards the versions of the dependencies (packages B and C).
Once package B is removed from A's dependencies (in February), we remove the edge A~$\to$~B in the dependency graph.
It is important to note that, by not accounting for versions in the nodes, this dependency graph is different from the dependency graph that \npm resolves to install new package versions when running the \texttt{npm~install} command~\cite{npm_install:online}.

\mainpoint{Extracting Dependency Change Events}
To build the \npm dependency graph, we need to extract and process all events that changed dependencies for all \npm packages.
In our study, we need to process two types of dependency change events for all \npm packages: 1) the addition of a new package dependency and 2) the removal a package dependency.
Since our dependency graph does not consider the package versions in their nodes, there is no need to account for events updating a package dependency version.
We use the \npm registry database to extract all the package dependency change events.
The \npm registry keeps a copy of the \texttt{package.json} file of all \npm packages in its database for all package versions.
The \texttt{package.json} file includes the list of maintainers, package description, keyword, license, the address of the source code repository, and the list of package dependencies.
The registry stores each package as a document that contains its metadata.

The \npm registry is powered by an Apache CouchDB database, which has a feature to set up a continuous stream of its data~\cite{npm_registry:online}.
The feature is typically used to set up continuous replication from the registry database.
We utilize this feature to retrieve a stream of all documents from the \npm registry (Step~\circled{1}).
For each document in the stream, we filter out the irrelevant documents (e.g., design documents) and for each package we collected the \texttt{package.json} file for each of its versions.

When we build the monthly dependency graph, we only use the most recent version of each package version to create our dependency graph. Hence we order every package release by its release date.
However, not all releases represent the stage of the package project at the target time.
Backports are commonly employed by package maintainers to fix older releases, and they could include old dependencies that no longer appear in the package's latest releases.
Hence, we filter out any release with a lower semantic version than its predecessor in relation to their respective release date (Step~\circled{2}).
For example, package A has released the version \texttt{3.6.0} in March 2020, but released a backport fix \texttt{2.1.0} in April 2020. Because the version \texttt{2.1.0} is smaller than version \texttt{3.6.0}, we disregard the version \texttt{2.1.0} in our analysis.

Finally, we extract the changes in the list of dependencies between versions (Step~\circled{3}).
We represent each change in the dependencies list as a dependency change event, which can be either an {add} or a {remove} event.
When a package releases its first version, we consider each of the dependencies required by that version as an {add} dependency change event.
In the following versions, we compare the list of dependencies on each version with the list in the previous version. If the dependency is absent in the newer version, we consider it to be a {remove} event; conversely, if the dependency is absent in the older version, we consider it an {add} event.

\mainpoint{Computing Monthly Centrality}
To obtain monthly snapshots of the centrality trends, we compute the centrality for the packages in the \npm ecosystem each month. Consequently, we need to build a dependency graph at each month of analysis.
Building separate graphs from scratch for every month can be an expensive operation and unpractical option, particularly for \npm, which contains more than a million packages.
To address this, we build the dependency graphs incrementally using the {add} and {remove} dependency change events that we explained previously.

In this study, we are interested in investigating the package centrality trends since the creation of the \npm ecosystem.
In particular, we study the period from December 2010 to December 2020.
To do so, we build the first graph up to December 2010 and calculate the centrality for every package in that graph (Step~\circled{4}).
Then, for each month, we update the graph snapshot to reflect the monthly changes in the ecosystem.
In total, we build 121 different versions of the dependency graph for the \npm ecosystem, one for each month between December~2010 and December~2020.

We use the monthly dependency graphs to compute the centrality of packages in the \npm ecosystem (Step~\circled{5}).
In order to compute the centrality, we use the Google PageRank algorithm~\cite{Brin_PageRank1998,Page_InfoLab1999}.
The algorithm is commonly used to rank software artifacts, e.g., JavaScript packages \cite{Wittern_MSR2016,npmrank:online} and Java components \cite{Inoue_TSE2005}.
The PageRank algorithm is a variant of the Eigenvector Centrality metric, which measures the importance of each node within the graph based on the number of incoming edges and the importance of the corresponding source nodes. The underlying assumption of PageRank is that a node is only as important as the nodes that link to it~\cite{Gleich_SIAM2015,Manaskasemsak_ICPADS2005}.
In our study and through the use of PageRank, the package centrality score is affected by both the number of dependent packages and the score of the dependent packages themselves.
Thus, packages obtain higher scores if their dependent packages themselves have high scores.

However, the centrality value of nodes in PageRank decays over time as the network grows~\cite{Berberich:07:PageRank}. This may impact the evolution analysis and means it is not meaningful to compare centrality values of packages on different periods as these will always tend to decrease (at least for growing networks).
To address this, we focus instead on analyzing the ranking of the nodes' centrality.
Once we compute the centrality for all packages on a particular month, we rank the packages based on their centrality values ($v_1, v_2, ..., v_n$) where $v_1$ is the most central package and $v_n$ is the least central package similar to prior work~\cite{Wittern_MSR2016}.
Finally, we invert the ranking in negative values ($-1 \times n$) to give a higher ranking value to the more central packages, and make the centrality ranking comparable to other metrics (e.g. downloads), where a higher value means higher importance.
With this, we have the centrality ranking position evolution for each package in the \npm ecosystem since its creation up until December 2020.

\subsection{Detecting Packages In Decline}
\label{sub:detect_decline}

Now that we have the evolution of all packages' centrality rankings, we use it to provide a reliable method to identify packages \decline.
To classify a package as \decline, we use its centrality trend of the latest six months.
We fit a linear function using the least-squares regression~\cite{Handbook:online}, then we analyze the slope~($m$) of the trend to identify a package \decline.
In our study, a package is classified \decline if its centrality trend shows a significant negative slope:

\begin{enumerate}
    \item \textbf{Slope:} the slope of the centrality trend for the last six months should be $m < v$, with default $v = 0$.
    \item \textbf{P-Value:} to test whether the negative slope is statistically significant, we perform the Wald Test with a conservative p-value ($p$) threshold, i.e., $p < \alpha$, with default $\alpha = 0.001$~\cite{Judge_Book1985}.
          The Wald Test is a way of testing the significance of particular explanatory variables in a statistical model.
\end{enumerate}

\begin{figure}[t]
    \centering
    \includegraphics[trim=0 0 0 .42in, clip, width=.85\linewidth]{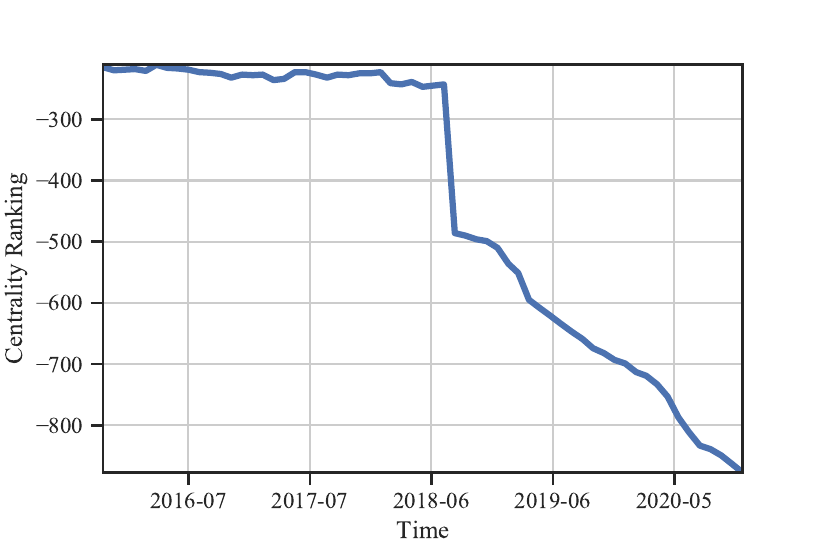}
    \caption{Example of a package trend \decline.}
    \label{fig:centrality_decline_example}
\end{figure}

In practice, our approach classifies packages as \decline when they have consistently fallen down in the \npm centrality rankings for six months.
\Cref{fig:centrality_decline_example} shows an example of the package istanbul-api, which is classified as \decline, with a clear decrease in the centrality rankings starting from mid 2018.
This decline can be justified by the incompatibility of the package with new Javascript features~\cite{istanbuljs_issues_321:online}, which led to the deprecation of the package later in April 2019~\cite{istanbuljs_issues_378:online}.

\section{Evaluation Datasets}
\label{sec:centrality:dataset}

To obtain a baseline for our approach, we devise a dataset containing packages \decline and packages \stable, so we can evaluate if our approach can reliably report packages \decline.
Unfortunately, there is no existing large dataset that captures the shifts in community interest we aim to evaluate.

To compensate for the absence of this ideal dataset, we build three different baseline datasets.
First, we build a corpus using metrics from the official search engine of \npm (\npms) to evaluate if centrality can detect packages \decline before \npms~(\Cref{sub:data_npms}).
Second, we collect data from the largest survey of the JavaScript community conducted by~\citet{StateOfJS:online}, which asked the opinion of more than 20 thousand developers about 20-30 popular \npm packages~(\Cref{sub:data_survey}).
With this baseline we aim to evaluate if centrality can capture the satisfaction/dissatisfaction of developers using the trend in centrality right before the survey took place.
Third, we craft a dataset of deprecated packages to evaluate if our approach can help identify the decline in popularity well before the maintainers deprecated the packages~(\Cref{sub:data_deprecated}).

\subsection{Extracting \npms Validation Baseline Corpora}
\label{sub:data_npms}

One of the most reliable platforms developers use to select \npm packages for their projects is the official \npm search engine, \npms~\cite{Abdellatif_IST2020,npms_about:online}.
The \npms engine continuously analyzes the \npm ecosystem, and collects 27 package metrics from different sources (e.g. package repositories on GitHub).
Using the collected metrics, a final score for each package is calculated based on three different aspects i.e., quality, popularity, and maintenance~\cite{npms_about:online,Abdellatif_IST2020}.
The higher the score of a package, the more popular, better quality and better maintained the community perceives the package to be.
Hence, a steep decline in a package score can be used as an indicator of a package \decline and a stable or increasing score can be an indicator of a package \stable.

\begin{figure}
    \centering
    \includegraphics[width=\linewidth, trim=12cm 6.7cm 12cm 7.5cm, clip]{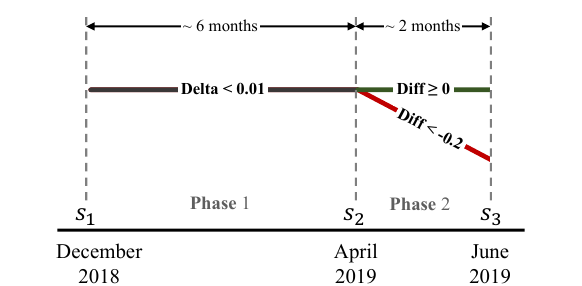}
    \caption{Timeline used to select validation baseline from \npms.}
    \label{fig:npms_snapshots_timeline}
\end{figure}

It is important to note however that we want to evaluate the hypothesis that \emph{centrality can better identify shifts in community interest than currently used metrics}.
Hence, we want to craft a dataset that allows us to use \npms score as validation, but is not directly influenced by the \npms score.
To this aim, we craft the dataset using a multi-phase approach, as illustrated in \Cref{fig:npms_snapshots_timeline}.
We first select packages that have shown a stable \npms score during a period (Phase 1), and use this same period to evaluate the centrality of a package.
Because the score metric is stable during this period, one would not be able to classify the packages \decline from \stable just by analyzing \npms score, and we can be sure centrality is not influenced by already reported metrics.
Then, in the subsequent period (Phase 2), we label packages \decline as the ones that have experienced a sharp decline in the \npms score and label packages \stable as the ones that have either remained stable or increased their \npms score.
Using this process to craft the baseline, we also have a starting date for packages \decline given by the \npms score, which is at the earliest the start of Period 2.
We can use this point in time, to evaluate how much in advance (if any) our approach can detect that a package is \decline before the decline is shown in the \npms score.

One limitation of using the \npms metrics is that \npms does not store the historical values of its packages' score.
We cannot pick any period interval for Phase 1 and Phase 2 and are limited by the snapshots of the entire \npms score ranking we collected in the past.
We collected \npms packages' scores on December 2018, April 2019, and June 2019 and we use the period of December 2018 to April 2019 as Phase 1, and use the period of April 2019 to June 2019 as Phase 2 of our dataset baseline.

All \npms scores vary from 0 (very low) to 1 (perfect score).
We start crafting our dataset by selecting packages that have a score 0.7 or higher, to prevent our analysis from focusing on very low-quality packages.
As we showed in~\Cref{fig:npms_snapshots_timeline}, in Phase 1 we consider all packages that have a score variation smaller than 0.01, which indicates to be relatively stable.
Then, we label as packages \decline, all packages that have exhibited a negative change in the \npms score between $S_2$ and $S_3$ by more than 0.2 score points.
We label as packages \stable, all packages that have exhibited the same score or higher between $S_2$ and $S_3$.
At the end of this process, this dataset contains a total of 4,457 packages, with 2,259 being labeled as \decline and 2,198 labeled as \stable.

The three thresholds used in the above methodology were determined as follows: the first threshold of 0.01 is the tolerance in the \npms score deltas in Phase 1. This threshold equals the mean value of changes in the \npms score between $S_2$ and $S_3$ and it is small enough to guarantee that package scores are stable for at least 6 months before April 2019.
The second threshold~0.2 is the minimum decrease in the \npms score to label a package as \decline. This threshold is equal to the value of standard deviation over the \npms scores and it is large enough to capture the significant score changes.
The last threshold, 0.7, is the minimum \npms score for a package on $S_3$ to be considered in our baseline dataset. This will minimize the risk of mislabeling our baseline by including low-quality packages with very low npms scores and it is a good compromise between the dataset size and quality.

\subsection{Survey Validation Baseline Corpus}
\label{sub:data_survey}

We want to evaluate if our approach can capture the shifts on the interest and satisfaction of the \npm community with popular packages.
While we cannot craft a dataset that reliably captures the \npm community interest without surveying a very large sample of JavaScript developers, we opted to use the data from the largest survey available on the JavaScript ecosystem: the State of JavaScript survey~\cite{StateOfJS:online}. %

The State of JavaScript survey is an extensive survey conducted by~\citet{StateOfJS:online} to assess the JavaScript community's views.
In 2019, the survey had a total of 21,717 respondents all across the globe~\cite{StateOfJS2019:online}.
The survey's primary focus is to ask JavaScript developers their opinion on a set of popular \npm packages.
Then, the survey ranks each package according to four categories:

\begin{enumerate}
    \item \textbf{Awareness:} share of total respondents that reported to have heard about the package. This category includes both developers who have experience using the package and developers never use the package before.
    \item \textbf{Usage:} share of total respondents that have used the package in their projects. This category does not consider if the developer is satisfied with using the package.
    \item \textbf{Interest:} share of respondents who did not use the package but are interested in using it in the future.
    \item \textbf{Satisfaction:} share of respondents that have used the package in the past and will continue to use it.
\end{enumerate}

To use the survey results, we use its GraphQL API\footnote{https://graphiql.stateofjs.com} to retrieve the summary of the responses for each package.

\subsection{Deprecated Packages Corpus}
\label{sub:data_deprecated}

With this third corpus, we want to evaluate if our approach can help identify packages \decline that have eventually been deprecated by maintainers.
Deprecated packages should not be reused by other packages or JavaScript applications and \npm warns developers when they install deprecated packages.
The goal of our analysis is to evaluate if centrality trends can capture the decline in the community interest well before the package is flagged as deprecated, which can help developers to migrate from these packages while they are still being maintained.

To craft this dataset, we need to collect a list of deprecated packages from the \npm ecosystem.
Similar to \Cref{sec:centrality:approach}, we started by retrieving the metadata for all packages from the \npm registry.
Then we capture metadata for packages with a deprecation message, which left us with a list of 44,857 packages.
However, developers use the \npm deprecation feature for various reasons, including renaming or merging packages. The following quote is an example of a deprecation message for a package whose maintainers used the depreciation feature to change the package name.
\saybox{Jade has been renamed to pug, please install the latest version of pug instead of jade~\cite{jade_deprecation:online}.}

To create a valid list of deprecated packages, we select the top 1,000 deprecated packages based on their \npms score on June 16th, 2019.
Then we manually classify packages to filter out cases where they are not an actual deprecation. For this aim, first, we verify if the deprecation note  discloses clearly that a package is actually deprecated.
If the deprecation message is not clear, we check the project status from the package's readme file, then the repository's readme file.
If needed, we follow relevant links in the deprecation messages or the readme files to remove ambiguity.
Finally, if the deprecation message mentions another package's name, we check if both are pointing to the same repository; if so, we examine the repository and its history to classify the case as a rename or not.
After applying our manual classification process, we find that only 556 out of the 1,000 packages are actual package deprecation cases. We use these 556 packages in our analysis later in the study.

\section{Results}
\label{sec:centrality:results}
This section describes our research questions. For each research question, we explain its motivation, illustrate our approach to answer the question, and discuss the findings.

\labelname{RQ1}\label{rq:one:num}
\subsection*{RQ1: How effective is our approach in detecting packages that are \decline?}
\label{rq:one}

\mainpoint{Motivation}
In this question, we investigate the performance of our approach of using the centrality trend to identify packages \decline.
The decline of package centrality could be a symptom that better alternatives have emerged or a shift happened in the community interest.
In the scientific literature, centrality has been used in many disciplines such as social networks to identify the central node of a network (e.g.,~\cite{Cadini,Maharani,Stergiopoulos,Hong}) and software engineering to understand the significance of software components (e.g., \cite{Wittern_MSR2016, Inoue_TSE2005}).
If the approach can aptly capture packages \decline, it can be embedded in package search engines, such as \npms, to increase developers' awareness of the community interest and help them make a better-informed decision to select or reevaluate their package dependencies.

\mainpoint{Approach}
We craft a baseline as described in \Cref{sub:data_npms} to evaluate our approach as a binary classification problem.
Then we use our approach to classify packages into two classes: \decline and \stable.

As mentioned in \Cref{sub:data_npms}, packages labeled \decline are packages that have experienced a sharp decline in the \npms ranking in a short period of two months, i.e., between $S_2$ and $S_3$.
We calculate the centrality in the last six months before $S_2$, when the packages were still stable in the \npms rankings.
This ensures that we evaluate if the centrality can be used as an early detector of packages \decline that only later will be observed in the \npms rankings.
Then, as described in \Cref{sub:detect_decline}, we classify packages that have a negative centrality trend slope (i.e.,~$m < v$ with default $v=0$) as \decline and other packages as \stable.

To evaluate the performance of our approach in identifying packages \decline, we report the well-know performance measures: precision ($P)$, recall ($R$), and $F_1$ score.
In the context of our evaluation, precision is the percentage of packages classified as \decline that are actually \decline (i.e.,~$Precision = \frac{T_p}{T_p+F_p}$), where $T_p$ is the number of packages labeled as \decline that are correctly classified as \decline; $F_p$ denotes the number of \stable packages classified as \decline.
Recall is the percentage of packages that correctly classified as \decline relative to all of the packages that are labeled as \decline (i.e.,~$Recall = \frac{T_p}{T_p + F_n}$), where $F_n$ measure the number of packages \decline that classified as \stable.
We then combine both precision and recall using the well-known $F_1$ score (i.e.,~$F_1 = 2\times\frac{P \times R}{P + R}$).

In addition, to mitigate the limitation of choosing a fixed slope threshold (i.e.,~$v = 0$) when calculating precision and recall, we also present the Area Under the Receiver Operating Characteristic Curve (ROC-AUC) value.
ROC-AUC is computed by measuring the area under the curve that plots the $T_p$ rate against the $F_p$ rate while varying the slope threshold used to determine if the approach should classify a package as \decline or not.
The ROC-AUC's main merit is that it reports the performance independently from the used threshold; it is also robust toward imbalanced data since its value is obtained by varying the classification threshold over all possible values~\cite{NamASE2015,LessmannTSE2008}.
The ROC-AUC has a value that ranges between 0 and 1, where a higher ROC-AUC value indicates better classification performance.

\begin{table}
	\centering
	\caption{Results of using the centrality trend to classify packages from the \npms validation baseline.}
	\label{tab:npmsdelta-classification-results}
	\begin{tabular}{l l r}
    \toprule

    \multirow{3}{*}{Dataset}
     & Total cases          & 4,457 \\
     & In decline           & 2,259 \\
     & Not in decline       & 2,198 \\
    \midrule

    \multirow{5}{*}{Performance}
     & True Positive ($T_p$)  & 1,969 \\
     & False Positive ($F_p$) & 498   \\
     & Precision ($P$)        & 0.80  \\
     & Recall ($R$ )          & 0.87  \\
     & $F_1$ score          & 0.83  \\
     & ROC-AUC              & 0.90  \\

    \bottomrule
\end{tabular}
\end{table}

\begin{figure}[t]
	\centering
	\includegraphics[trim=0 0 0 .42in, clip, width=.85\linewidth]{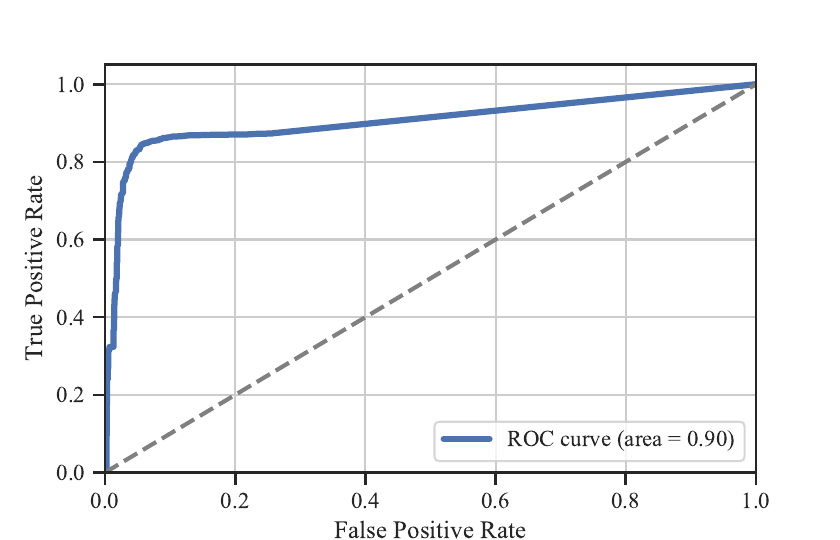}
	\caption{ROC curve with the AUC value for the evaluation based on the \npms baseline.}
	\label{fig:npms_roc_auc}
\end{figure}

\mainpoint{Results}
As shown in \Cref{tab:npmsdelta-classification-results}, we evaluate our approach on 4,457 \npm packages where 2,259 are labeled as \decline and 2,198 are labeled as \stable.
The results show that our approach of using the centrality trends correctly identifies 87\% of the packages \decline with a precision equal to 0.80.
That is, for every five packages classified as \decline, four were correctly classified and one was wrongly flagged as \decline.
This indicates that our approach can aptly identify packages \decline before they are actually shown in the \npms rankings, with an $F_1$ score of 0.83 and ROC-AUC of 0.90 As shown in \Cref{fig:npms_roc_auc}.
The figure shows the false positive rate on the x-axis and the true positive rate on the y-axis, while the solid line represents the value of each of them based on a range of possible thresholds.

\begin{figure}[t]
	\centering
	\begin{subfigure}[t]{0.48\linewidth}
		\centering
		\includegraphics[trim=0 0 .5cm 0cm, clip,width=\linewidth]{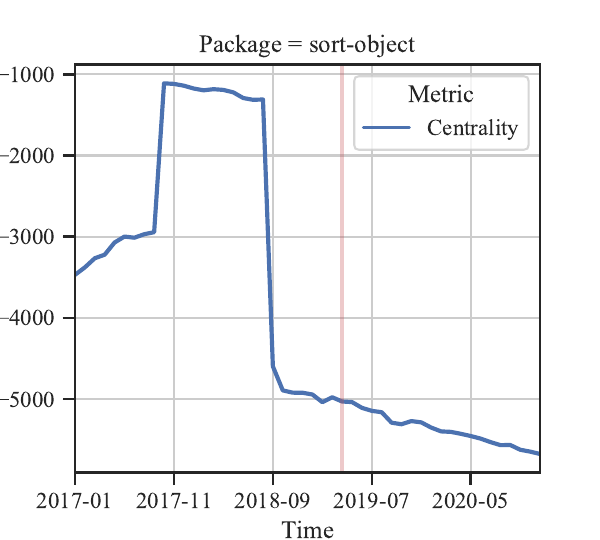}
		\label{fig:late_detection_sort-object}
	\end{subfigure}
	~
	\begin{subfigure}[t]{0.48\linewidth}
		\centering
		\includegraphics[trim=0 0 .5cm 0, clip,width=\linewidth]{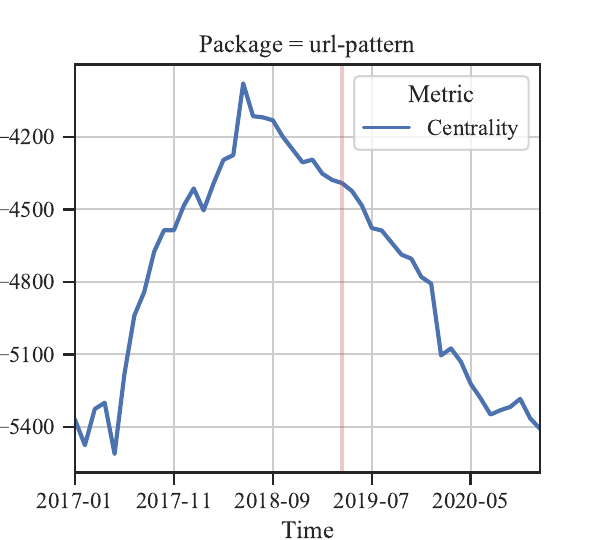}
		\label{fig:late_detection_url-pattern}
	\end{subfigure}

	\fullcaption[Examples of packages that our approach only detected the decline after the \npms score.]{
		The red vertical line indicates the time of $S_2$.}
	\label{fig:late_detection}
\end{figure}

We analyze the 290 packages that were \decline, but where our approach could not identify their decline using centrality.
Out of the 290 cases, 217 (74.83\%) packages exhibited a centrality decrease only after April 2019 ($S_2$), showing that in these cases the \npms metrics decrease before the centrality.
\Cref{fig:late_detection} shows examples of packages that our approach could not detect packages \decline in advance of \npms.
In the figure, both packages show a decrease in centrality before April 2019 ($S_2$). However, our approach requires a statistically significant decrease over a six months period, with a very conservative default threshold $\alpha = 0.001$ to detect the packages as \decline.
Hence, our approach detected the packages as \decline after $S_2$, when the decline became statistically significant.

We also examine the 498 packages that were \stable, but where our approach wrongly identifies them as \decline.
We observe that out of the 498 cases, 384 (77.11\%) packages have less than 100 dependent packages.
In the rest of the false positive cases, we observe that most of the packages (112 out of 114) have their number of dependents increasing, however their centrality is decreasing.
For example, the package mongoose is a popular object modeling tool whose dependents increased from 4,995 to 5,568, whereas its centrality ranking declined from -443 to -484.
The main factors behind these false positive cases can be explained by the following:
\begin{enumerate}
	\item The dynamic of the centrality ranking tends to punish packages that do not gain more dependents (directly or indirectly) on them compared to other packages in the same ranking tier.
	      In the mongoose example, even with the 11.47\% increase in the number of dependents, the number of dependents and their centrality were not enough to maintain the centrality ranking compared to other packages in the same ranking tier.
	\item In packages with a small number of dependents, the centrality trend can be affected by a small number of community members that do not reflect the overall community interest. This could explain 52 (17.93\%) of the 498 false negative cases and 384 (77.11\%) of the 498 false positive cases.
\end{enumerate}

\mainpoint{Impact of Moving Averages}
Simple moving averages (SMA) is a technique used to reduce the noise in the time-series data~\cite{James_Cambridge1068}.
In this RQ, we use the trend of the monthly centrality rankings to detect packages in decline. However, using the SMA to smoothen the trends may result in improving the performance of our approach.
In our context, we experiment using the technique to reduce the effect of noise in the monthly centrality data. To do so, we re-run our experiments on the \npms validation baseline. For each package in the baseline, we compute the simple moving averages (based on 4 months average) for its monthly centrality rankings. Next, we apply our approach in detecting the centrality decline on the SMA values. 
The result of the experiment shows that incorporating the moving averages improved the precision of our approach from 0.80 to 0.85. However, it slightly decreases our approach’s recall from 0.87 to 0.83, while keeping the F1-score almost constant (from 0.83 to 0.84).
Finally, since using the moving averages requires more extended history, the number of packages that our approach can be applied on is reduced slightly from 4,457 to 4,272 packages.

\conclusion{
	The result shows that our approach can correctly detect~\textbf{87\%} of packages \decline with a precision of 0.80, an F1-score of 0.83 and an ROC-AUC of 0.90.
}

\labelname{RQ2}\label{rq:two:num}
\subsection*{RQ2: How early can our approach detect packages that are \decline?}
\label{rq:two}

\mainpoint{Motivation}
Once we learned that our approach is effective in identifying packages \decline, we would like to know how early in advance can our approach detect the decline.
Identifying packages \decline as early as possible is essential for taking proactive action to mitigate the decline of the package.
Also, it increases the awareness of the community about possible better alternatives by allowing developers to avoid selecting declining packages and to pay more attention to the alternatives that are increasing in centrality.
Package maintainers can also use our approach as a sign of a decrease in community interest in their package, which can motivate them to remediate possible causes of dissatisfaction or make them focus on other solutions altogether.
Furthermore, developers that reuse packages can use the centrality trend as an early indicator of decline to look for alternatives long before their dependencies become unmaintained.

\mainpoint{Approach}
To evaluate how early our approach can detect packages \decline, we employ a sliding window technique.
Since we calculate centrality at the granularity of months, we slide the analysis window back in time, sliding our window of six months one month at a time.
We recalculate the \decline analysis after each window sliding (i.e., month)
by applying the same method explained in \Cref{sub:detect_decline}.
We continue this process as long as the \decline analysis continues to identify the package as \decline.

Note that since we use a 6-month window to detect the deceline,
when we report that our approach captured a package \decline 4 months in advance, this means that the slope of the centrality trend consistently decreased in the 6 months prior to these 4 months.
That is, the package is exhibiting a decrease in the centrality rankings for up to 10 months.

We used our three different dataset baselines to evaluate how early our approach can detect packages \decline.
We evaluate how early our approach can detect packages \decline based on all packages that our approach classifies as packages \decline.

\begin{enumerate}
    \item  \textbf{\npms dataset:} This dataset was crafted from the \npms rankings, as explained in \Cref{sub:data_npms}.
          In this dataset, we start measuring the packages \decline before April 18th, 2019, where the \npms score was still stable.

    \item \textbf{Deprecated dataset:} This dataset was crafted from the deprecated \npm packages, as explained in \Cref{sub:data_deprecated}.
          In this evaluation, we aim to assess how far in advance we can use our approach to identify packages that have become deprecated.
          In this evaluation, we use the deprecation date of each package as the starting point to measure whether the package is \decline.

    \item \textbf{State of JavaScript dataset:} We collect this dataset from the State of JavaScript survey of 2019~\cite{StateOfJS2019:online}, as explained in \Cref{sub:data_survey}. We label packages with a share of satisfaction less than 50\% as \decline and the rest of the packages as \stable. In this evaluation, we measure the packages \decline before November 25th, 2019, which is the date of receiving the first survey response.

\end{enumerate}

\begin{table}
    \centering
    \caption{Results of three datasets on how early in months our approach can detect packages \decline.}
    \label{tab:how_early_results}
    \begin{tabular}{lrr|rr}
    \toprule
    \multirow{2}{*}{Dataset} & \multirow{2}{1.3cm}{Labeled as \decline} & \multirow{2}{1.45cm}{Classified as \decline} & \multicolumn{2}{c}{\textbf{Time (months)}}          \\
                             &                                          &                                              & Mean                                       & Median \\

    \midrule
    \npms                    & 2259                                     & 2467                                         & 18.35                                      & 12.57  \\
    Deprecated               & 552                                      & 446                                          & 16.15                                      & 13.29  \\
    Survey                   & 4                                        & 3                                            & 13.13                                      & 4.80   \\
    \bottomrule
\end{tabular}

\end{table}
\mainpoint{Results}
\Cref{tab:how_early_results} presents the results of our experiment, showing how far in advance our approach can detect packages \decline.
The first row in the table shows that our approach classifies 2,467 packages from the \npms dataset as \decline with an average of 18.35 (median~=~12.57) months before April 2019 ($S_2$).
To reiterate, only after the $S_2$ date, these packages have shown a steep decline in the \npms scores.
Our results show that half of the packages were experiencing consistent centrality decline for more than a year before this decline was captured by the \npms metrics.

The second row in \Cref{tab:how_early_results} shows the results for the deprecated dataset.
Our approach was able to identify the centrality decline on average more than a year (16.15 months) before the packages became deprecated.
Also, the decrease in the centrality rankings captures the decline of 446 out of 552 deprecated packages.
Our results indicate that the centrality trend can be used as an early indicator of deprecated packages, with a good recall, capturing 80\% of the deprecated packages.

Finally, the third row in \Cref{tab:how_early_results} shows the results of our evaluation using the State of JavaScript survey dataset.
Our approach correctly classified three out of the four labeled \decline packages with an average of 13.13 (median~=~4.80) months before the first survey response date without any false alarms.%

\begin{figure}[t]
    \centering
    \includegraphics[trim=0 0 0 .42in, clip,width=.85\linewidth]{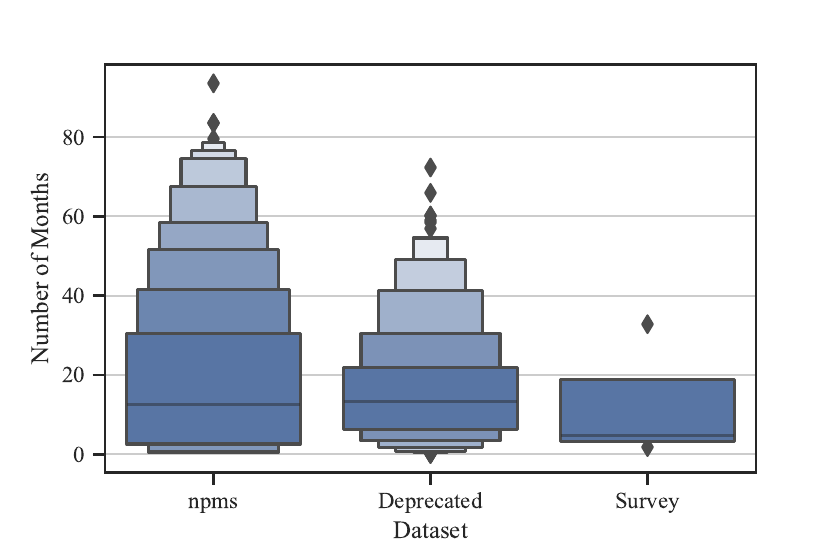}
    \caption{Letter-value plots for the distribution of how early our approach can detect packages \decline.}
    \label{fig:how_early_distribution}
\end{figure}

\Cref{fig:how_early_distribution} shows the distribution of time in months for how early our approach can detect packages \decline across the three datasets.
The figure shows that our approach detects 25\% of packages \decline more than 31 months before the significant \npms score decrease, and 22 months before a package got deprecated.

\conclusion{
    The results show that our approach can detect packages \decline on average \textbf{18.35 months} before the \npms score declines. Also, it detects packages \decline on average \textbf{16.15 months} before a package gets deprecated.
}

\labelname{RQ3}\label{rq:three:num}
\subsection*{RQ3: How does our approach compare to other metrics in detecting packages that are \decline?}
\label{rq:three}
\begin{figure*}[t]
	\centering
	\includegraphics[trim=0 0 0 .1in, clip,width=\linewidth]{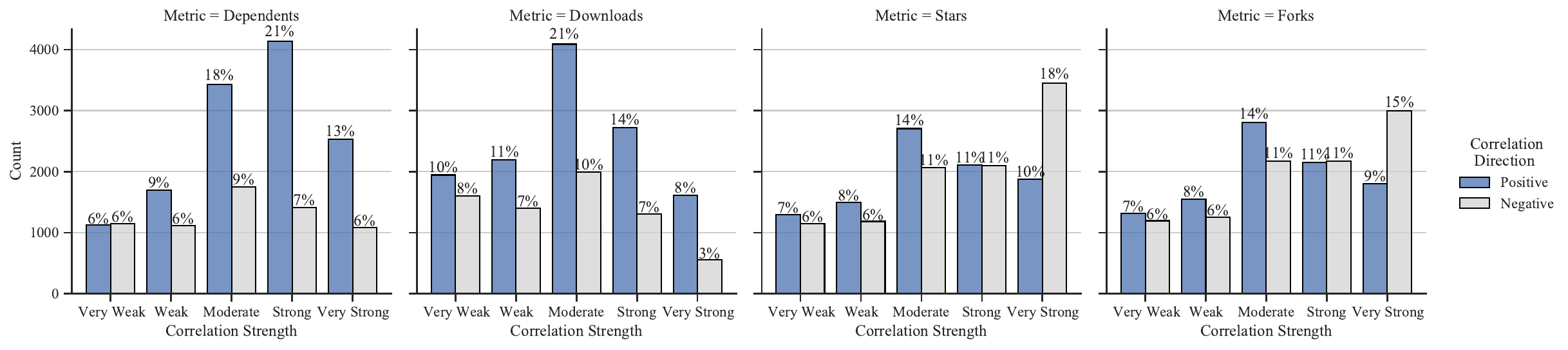}
	\caption{The distribution of the correlation between centrality and the metrics.}
	\label{fig:displot_metric_correlations}
\end{figure*}

\mainpoint{Motivation}
After determining our approach's effectiveness in detecting packages \decline, months in advance, we would like to know if other widely used metrics already capture (or complement) the information centrality indicates.
There are already several metrics, e.g., as the number of GitHub stars from their repository project, that aim to provide a popularity indicator of \npm packages and have been used by prior work, (e.g.,~\cite{Abdellatif_IST2020,Borges_ICSME2016,Borges_JSS2018,Papamichail_QRS2016,Zhu_ESEM2014,Zerouali_SANER2019}).
If centrality is already properly captured by other widely used metrics, there is no incentive to incorporate centrality in the current package platforms.
If the centrality trends, however, provide a new perspective on the popularity and community interest of a package, there is a good motivation to make the centrality information more accessible to developers to improve their community awareness.

\mainpoint{Approach}
We are particularly interested in assessing how much of the centrality is already captured by metrics that the \npms analyzer uses. In particular, we studied metrics that present the number of dependents, number of package downloads, Github stars, and Github forks.
We evaluate if the centrality trend correlates with these metrics and whether we could use these previously used metrics to detect packages \decline, with similar or better performance than our centrality trends.

To compare our approach with the other metrics, we start by collecting the monthly number of dependents, downloads, stars, and forks of 40,619 packages in \npm.
We retrieve the number of monthly dependents using the dependency graphs we build to measure the centrality, explained in \Cref{sub:calculat_centrality}.
For the number of downloads, we use the \npm REST API\footnote{https://api.npmjs.org} to collect the daily number of downloads for the time between each package creation date (not before February 2015, which earliest data that the API keeps) until December 2020.
Then we aggregate the daily downloads for every month.

The GitHub API does not provide an endpoint to retrieve the historical number of stars and forks.
To overcome this challenge, we rely on the API of Porter.io,\footnote{https://porter.io} a service that analyzes Github continuously and retrieves the historical number of stars and forks for a wide range of repositories.
Thus, we use Porter.io's API to collect the historical number of Github stars and forks for package repositories with more than 100 stars in \npms at December 27th, 2020.
We omit packages with fewer than 100 stars, to prevent our analysis from being dominated by packages that are seldom used by the community.

After collecting the metrics for all packages with more than 100 stars, we notice that not all packages have sufficient data for our analysis.
For instance, some packages lack sufficient historical data or one or more of their metrics have all the data points as zero, e.g., packages that have no dependents.
Therefore, to simplify our analysis and report results from a uniform dataset, we exclude packages that do not have sufficient information for all metrics.
This step excluded 21,201 packages from the initial set of 40,619; thus, our analysis is based on 19,418 packages.

To evaluate if the other metrics' trends indicate the same trend as centrality rankings, we test the correlation between the monthly centrality trend and each of the other metrics' monthly trends.
We use Spearman's rank correlation test, and we apply the correlation test on the metrics for each package separately.
We use Spearman's rank correlation coefficient since our dataset is not normally distributed~\cite{kendall1938new}. Spearman's rank correlation coefficient ($\rho$) has a value that ranges between +1 and -1. In our context, +1 means that a metric value always increases when the centrality increases and -1 means that a metric value always decreases when the centrality increases. A Spearman ($\rho$) of zero indicates no correlation between the metic and the centrality~\cite{kendall1938new,Fowler_Book2009}.

\mainpoint{Results}
\Cref{fig:displot_metric_correlations} shows the distribution of Spearman's rank correlation coefficient ($\rho$) between the centrality trend and the trend of each of the other popular metrics. Following the guidelines of \citet{Fowler_Book2009}, we group the correlation distribution into five intervals: very weak correlation (0.00 to 0.19), weak correlation (0.20 to 0.39), moderate correlation (0.40 to 0.69), strong correlation (0.70 to 0.89) and very strong correlation (0.90 to 1.00).
The figure plots the correlation results for 19,418 packages. %
We observe that centrality and the evaluated metrics have correlations that spread all the spectrum from a perfect positive correlation ($\rho = +1$) to a perfect negative correlation ($\rho = -1$).
Overall, this shows that centrality is not aligned to the other metrics for most packages, indicating that centrality may provide new information that is not captured by the other metrics.
Next, we discuss the comparison to each metric and its implications.

\begin{figure}[t]
	\centering
	\begin{subfigure}[t]{0.48\linewidth}
		\centering
		\includegraphics[trim=0 0 .5cm 0cm, clip,width=\linewidth]{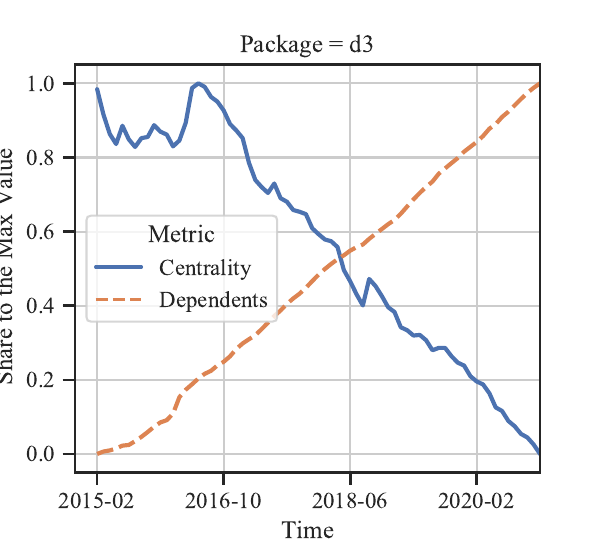}
		\label{fig:example_corr_negative_d3}
	\end{subfigure}
	~
	\begin{subfigure}[t]{0.48\linewidth}
		\centering
		\includegraphics[trim=0 0 .5cm 0, clip,width=\linewidth]{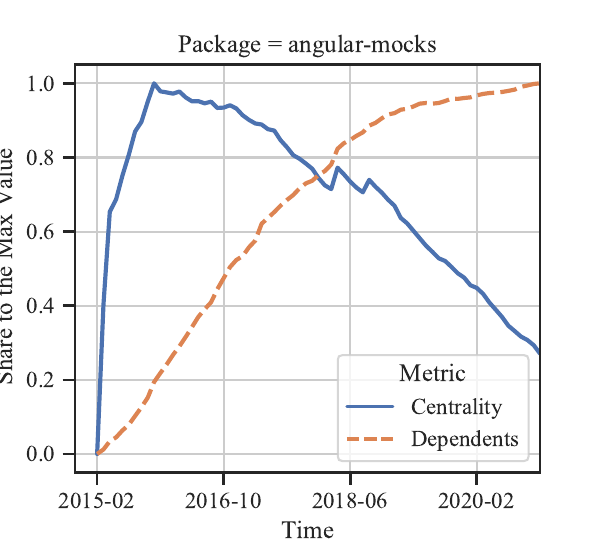}
		\label{fig:example_corr_negative_angular-mocks}
	\end{subfigure}

	\fullcaption[Examples of packages with strong negative correlation between the centrality trend and the number of dependents trend.]{
		We normalize metric values to range between 0 and 1.}
	\label{fig:example_corr_negative}
\end{figure}

As shown in \Cref{fig:displot_metric_correlations}, the dependents metric shows the strongest correlation with centrality amongst the evaluated metrics.
Roughly a third of the packages (34\%) have a strong or very strong correlation between its number of dependents and centrality.
This is somehow expected since packages with high centrality tend to have many dependents and vice versa.
Still, this strong correlation does not hold for the majority of packages that we evaluated because our approach to calculate the centrality uses an algorithm that considers not only the number of dependents but also the importance of each of them.
This explains why 13\% of the packages have a strong negative correlation between the number of dependents and centrality.
Such packages, such as the examples in \Cref{fig:example_corr_negative}, have shown a steady increase in the number of dependents but an equally steady decrease in their centrality in \npm.

\begin{figure*}[t]
	\centering
	\includegraphics[trim=0 0 0 .2in, clip, width=\linewidth]{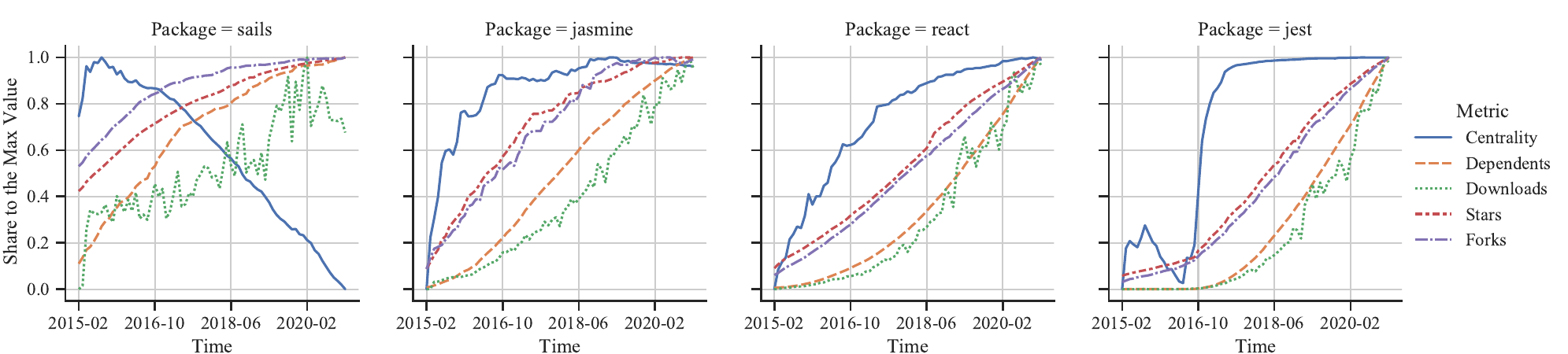}
	\fullcaption[Line plots showing the trend of centrality alongside with the trend of other metrics.]{
		We normalize metric values using the min-max method where values range from 0 and 1~\cite{Codecademy_Normalization:online}.
	}
	\label{fig:metrics_trends_examples}
\end{figure*}

The number of downloads also has a strong positive correlation with centrality in 22\% of the packages.
Similar to the case of the number of dependents, it is expected that packages that rise in the centrality ranking will have an increase in the number of downloads.
In 36\% of the cases, however, the centrality and the number of downloads are only weakly correlated (positively and negatively), and in 10\% of the packages, they have shown a strong negative correlation.
As shown in the \moment example (\Cref{sec:centrality:motivation}), these are the packages that, albeit having a constant increase in their downloads, are falling in the ranking and becoming less central in the \npm network.
These are the packages in which centrality can work best as an indicator of community interest.
The number of downloads depends on the number of installed systems, which may take a longer time to reflect the package's actual community interest.

The stars and forks metrics have approximately half the packages positively correlated with centrality and half the packages negatively correlated with centrality.
This is a consequence of the monotonic characteristic of stars and forks.
Projects tend to always increase their number of stars/forks, as contributors only rarely remove stars from a project.
In fact, in our dataset only 2.39\% of the packages showed a substantial decrease in the number of stars and forks in their life cycle.
Centrality, on the other hand, may increase and decrease as the community shifts its interest to the package or away from it.

To gain a better understanding of how these metrics are different, we use the following process to select four package examples:
1) We use the State of JavaScript 2019 survey that we explain in \Cref{sub:data_survey} to select popular packages.
2) The survey includes 28 \npm packages; we order them based on the community satisfaction score and select a package from each quartile, i.e., Sails.js, Jasmine, React, and Jest with satisfaction scores 26\%, 67\%, 89\%, and 96\%, respectively.
\Cref{fig:metrics_trends_examples} plots each of the four packages with their monthly trend for all metrics.

With the decrease in maintenance activities and the increase in the number of unfixed bugs, developers start discussing the quality and health of the package Sails.js~\cite{Sails_status_HackerNews:online,Sails_is_dying_HackerNews:online}.
We observe from \Cref{fig:metrics_trends_examples} that the package Sails.js has a decreasing centrality trend since 2015; however, all other metrics continued to increase. The centrality trend is more consistent with the survey results, where 74\% of the developers (1,166~developers)  that said they used the package Sails.js responded that they would not use the package again.
Even though the package decreases in centrality, the package is still increasing in the number of downloads and other metrics.

Conversely, the packages Jasmine and React, which have relatively higher satisfaction scores, show a consistent increase in the centrality trend.
The package Jest showed an interesting change in the centrality evolution. The package had known performance issues until mid 2016~\cite{Jest_performance:online}, where the centrality decreased.
After the maintainer of Jest performed a complete rewrite of the package to overcome its issues~\cite{Jest_rewreten:online}, and having these changes well-received by the community~\cite{Jest_HackerNews:online}, Jest started showing a significant increase in centrality.
By looking at \Cref{fig:metrics_trends_examples}, we see that only the centrality measure captures the changes of the community's interest toward Jest.

\conclusion{
	Centrality tends to provide trends that are different from those provided by other metrics such as dependents, downloads, stars, and forks.
}

\section{Tool Prototype}
\label{sec:centrality:implications}

\begin{figure*}[t]
    \centering
    \fbox{\includegraphics[trim=.5cm 0cm .5cm 0cm, clip, width=\linewidth]{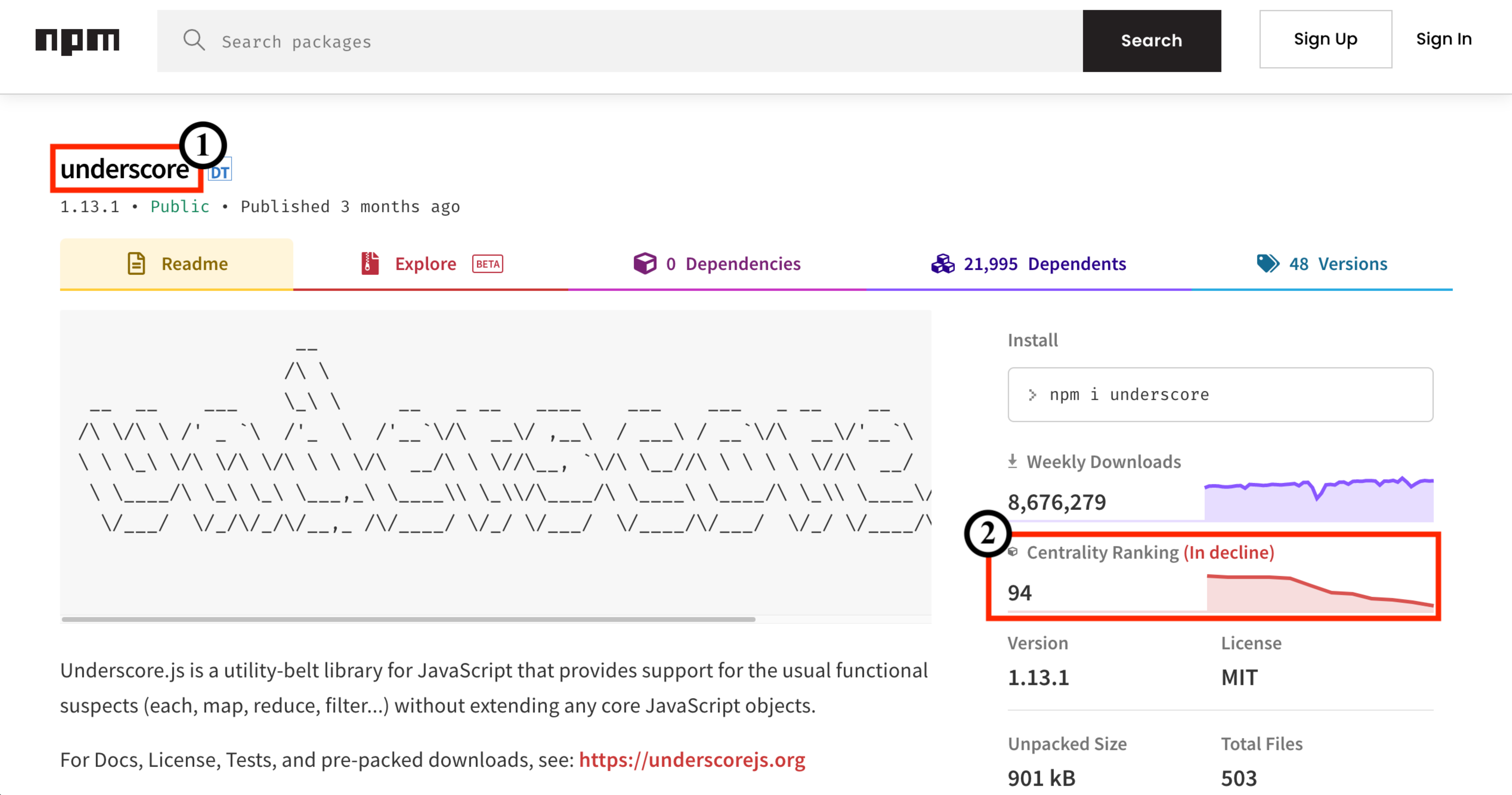}}
    \caption{A screenshot of the \npm website showing the package underscore with the integrated centrality information from our Chrome extension.}
    \label{fig:tool_screenshot}
\end{figure*}

The main implication of our study is that reporting the centrality trends of packages as a popularity metric in \npm can be very informative for developers.
Developers should use the centrality trend, together with other popularity metrics, to have a better informed assessment on which packages to select.
To enable this, we build a prototype web browser extension called \texttt{Centrality~Checker} that uses our approach of detecting package \decline. Our prototype extension helps inform developers about the centrality trend when they browse a package on the official \npm website.\footnote{https://www.npmjs.com/}

We build the tool as a Chrome Extension. Users can activate our extension in their Chrome browser. Once they browse a package on the \npms website, our extension includes the package centrality trends and the result of examining if the package is \decline into the \npm website.
The initial view when a user browses a package on \npm shows the centrality trend of the last year.
Users can hover over the centrality trend chart to explore the monthly centrality ranking values from the last year.
\Cref{fig:tool_screenshot} shows an example of an \npm package with the proposed Chrome extension enabled. In this example, we show the package underscore~\circled{1} with the centrality information embedded~\circled{2}.

When a user browses a package on the \npm website, the extension sends a request to a backend server to retrieve the needed data to render and embed the centrality ranking into the \npm website. The backend continuously retrieves the dependency change events from the \npm registry and calculate the centrality once every month as described in \Cref{sub:calculat_centrality}. The backend then determines whether each package is \decline using the approach described in \Cref{sub:detect_decline}. Finally, the backend caches the results to be served efficiently to our web browser extension.
The tool is publicly available and can be installed through the Chrome Web Store.\footnote{\url{https://chrome.google.com/webstore/detail/centrality-checker/bmpafkghbmojppjoeienibieljacdoaj}} Also, we open sourced the tool on Github.\footnote{\url{https://github.com/centrality-checker/chrome-extension}}

\begin{figure}[t]
    \centering
    \includegraphics[trim=0 0 0 .35in, clip, width=.85\linewidth]{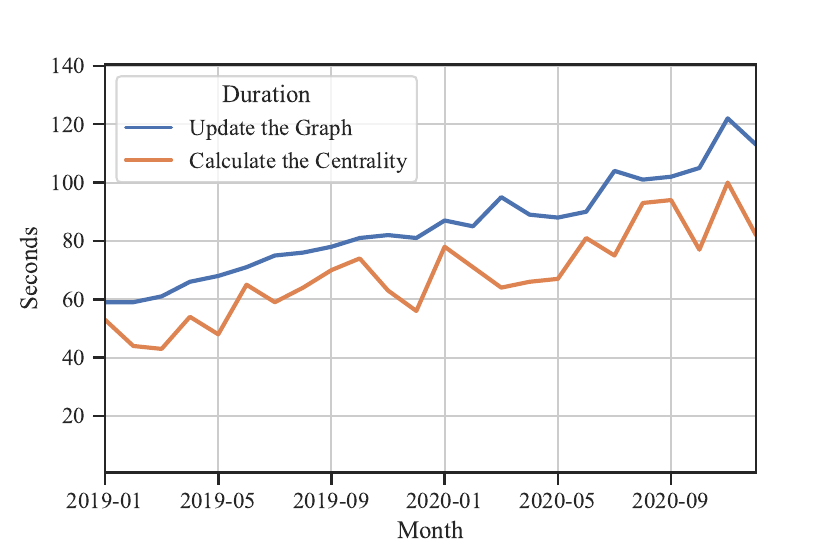}
    \fullcaption[The time required to update the dependency graph and calculate the centrality for all packages.]{The experiment was performed on a conventional machine with an Intel Core~i5 processor and 16GB of memory.}
    \label{fig:graph_scalability}
\end{figure}

\mainpoint{Scalability}
With the exponential growth in the number of packages in the \npm ecosystem~\cite{Decan_EMSE2019}, the time required to incrementally build the monthly dependency graph and calculate the centrality for all packages increases over time.
In particular, as shown in \Cref{fig:graph_scalability}, the time required to update the dependency graph increased from 1 minute in January 2019 to 2 minutes in December 2020.
The same goes for the time needed for calculating the centrality and detecting packages that are in decline, which increased from 50 seconds to 100 seconds.
However, even with this increase, the cost of running our approach is relatively low and it can scale to handle the rapid growth of the \npm ecosystem.

\section{Related Work}
\label{sec:centrality:related_work}
In this section, we discuss the work that is most related to our study.

Several studies examine the overall growth of software ecosystems.
For example, \citet{Wittern_MSR2016} did the first large-scale study of the \npm ecosystem. They study the evolution of the \npm ecosystem regarding growth and development activities. The study found that only 27.5\% of packages in the \npm ecosystem are depended upon, indicating that developers largely depend on a  core set of packages. %
\citet{Decan_EMSE2019} also empirically compare the evolution of the dependency network in seven software packaging ecosystems. Their results show how fast each packaging ecosystem and packaging dependency network is growing over time. They observe the continuing growth of the number of packages and their dependency relationships. %
Some other work also studies the evolution of software ecosystems~(e.g.,~\cite{Kikas_MSR2017,German_CSMR2013}).
In the same line with these existing studies, our work examines the evolution of \npm ecosystem in terms of its dependency graph. However, we focus on employing the \npm dependency graph and calculate the centrality for each package to identify \npm packages that are \decline.

Other work has been done to examine software projects that are not active anymore. For example, \citet{Coelho_ESEM2018} use machine learning classifiers to identify unmaintained GitHub projects. They also examine the level of maintenance activity of active GitHub projects, aiming to detect unmaintained projects. In an extension work,~\citet{Coelho_IST2020} developed a metric to alert developers about the risks of depending on a given GitHub project based on the built ML classifiers. In the context of the Python ecosystem, \citet{Valiev_FSE2018} studied the factors that affect the sustainability open source projects. Their results show that the centrality of a project in the ecosystem dependency network has a high impact on the project activities. Other works also investigate the overall popularity of open source projects. For example, \citet{Borges_ICSME2016} studied the popularity of GitHub repositories. They were able to identify four patterns of popularity growth, which relate to factors such as stars and forks.
As shown in the work mentioned above, examining the level of activity of an open source project is of critical importance, in particular, for packages in software ecosystems in order to maintain healthy dependencies.
Hence, our work addresses this issue by detecting which \npm packages are \decline.

There is also a body of research that investigates specific aspects of packages in a software ecosystem, including the source code size of packages~\cite{Abdalkareem_EMSE2020}, the impact of forks on the popularity of packages~\cite{Zhu_ESEM2014}, conflicts between used JavaScript packages~\cite{PatraICSE2018} or Python packages~\cite{wang2020watchman}, identifying breaking updates in \npm package~\cite{MujahidMSR2020}, and studying cross-project bugs that may impact a large part of a software ecosystem~\cite{ma2020impact}. Similar to these aforementioned studies, we focus on one aspect of the used packages in the \npm ecosystem: the package centrality. We propose the use of package centrality to identify packages \decline and evaluate its effectiveness in the \npm ecosystem.

\section{Threats to Validity}
\label{sec:centrality:threats_to_validity}

\noindent In this subsection, we discuss threats to the validity of our study.
\subsection{Threats to Internal Validity}

Threats to internal validity are related to experimenter bias and errors.
A limitation of our approach is that it only considers dependencies between packages in \npm.
This limitation will impact the centrality of packages that are not meant to be reused by other packages, but other JavaScript applications.
Future work should investigate how to incorporate JavaScript applications in the network and how to attribute their importance in the \npm network (e.g., using the number of stars in GitHub).
In our approach, the package importance is calculated by the centrality of its dependents, however, applications are not meant to be reused by other projects.

Another important threat to internal validity concerns the datasets that we used as baselines when evaluating our approach.
In our baseline datasets, we used various thresholds that impact which packages to include and their labeling.
Since having a gold standard for \npm's community interest is very difficult, we combine evaluations made from three datasets to mitigate for the lack of a large-scale ground truth.
Still, there is a need for a long term evaluation of the centrality as a complementary metric for current popularity metrics. Future work could investigate if developers find centrality a useful metric when selecting packages.
Finally, our approach may contain bugs that may have affected our results.
We made our scripts and dataset publicly available to be fully transparent and have the community help verify (and enhance) our approach~\cite{datasetonline}.

\subsection{Threats to External Validity}

Threats to external validity are related to the generalizability of our findings. Our investigation focused entirely on the \npm ecosystem, which has very particular characteristics: a centralized package registry, hundreds of thousands of software packages, and a very active and popular programming language.
Also, the size of packages in the \npm ecosystem is relatively small compared to the size of modules and software components in other ecosystems and programming languages. The small package size in the \npm ecosystem could lead to different dynamics compared to other ecosystems, which might significantly affect packages' characteristics such as the maintenance lifetime, release span, and barriers to migrate to other packages.
While centrality is a commonly employed metric to evaluate the importance in highly-connected systems, such as software ecosystems, the performance of our approach might be linked to the highly dynamic characteristics of \npm.
Future work needs to investigate if a similar approach can also help identify packages in decline in other ecosystems such as PyPi and Maven.

\section{Conclusion}
\label{sec:centrality:conclusion}

This paper presents a novel and scalable approach for using the centrality of packages to identify packages \decline.
We evaluate our approach in \npm, one of the largest and most popular software ecosystems.
Our evaluation showed that the centrality trends were effective at identifying packages \decline (RQ1).
When classifying packages as \decline and \stable, our approach can distinguish between the two classes with an AUC of 0.9.
Our approach correctly classified 87\% of the packages \decline, on average 18 months before the \npms aggregated score (RQ2).
By evaluating the correlation between centrality and current popularity metrics (e.g., number of downloads), we have shown that centrality trends can provide new information, not currently captured by \npms (RQ3).
We implemented our approach in a tool that can be used by developers to complement current \npms popularity metrics with our centrality trends.
Our approach can provide a more accurate depiction of the shifts the community interest makes and help inform developers when selecting packages for their software projects.

Our paper outlines some directions for future work. First, in this paper, we use centrality as an indicator of packages in decline. We believe investigating and understanding why packages' centrality is rising or declining is critical since it helps developers make more informed decisions. Another interesting followup work is to propose an automated approach to finding future central packages so they can receive the attention needed to boost their evolution as early as possible. Finally, after identifying packages in decline, the next step should be assisting developers in replacing them. Thus, we plan to develop an approach that suggests better alternative packages for those in decline.

\bibliographystyle{IEEEtranN}
\bibliography{bibliography.bib}

\begin{IEEEbiography}[{\includegraphics[width=1in,height=1.25in,clip,keepaspectratio]{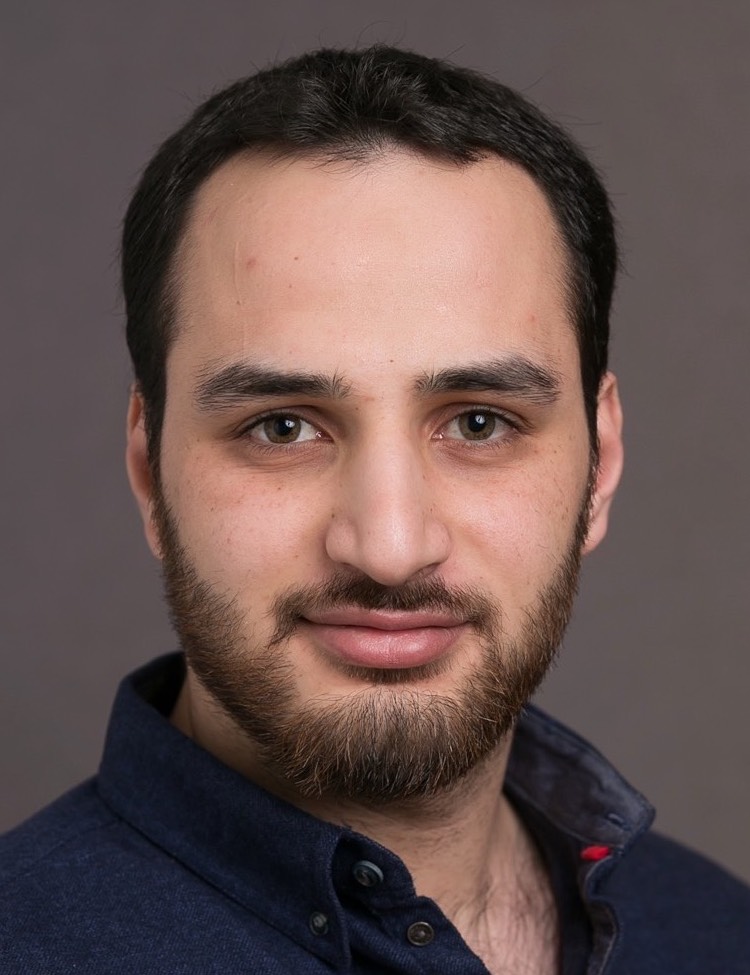}}]{Suhaib Mujahid}
  is a Ph.D. candidate in the Department of Computer Science and Software Engineering at Concordia University. He received his master’s in Software Engineering from Concordia University (Canada) in 2017, where his work focused on detection and mitigation of permission-related issues facing wearable app developers. He did his Bachelors in Information Systems at Palestine Polytechnic University.
  His research interests include software ecosystems, machine learning on code, software quality assurance, mining software repositories and empirical software engineering. You can find more about him at \url{https://suhaib.ca}.
\end{IEEEbiography}

\begin{IEEEbiography}[{\includegraphics[width=1in,height=1.25in,clip,keepaspectratio]{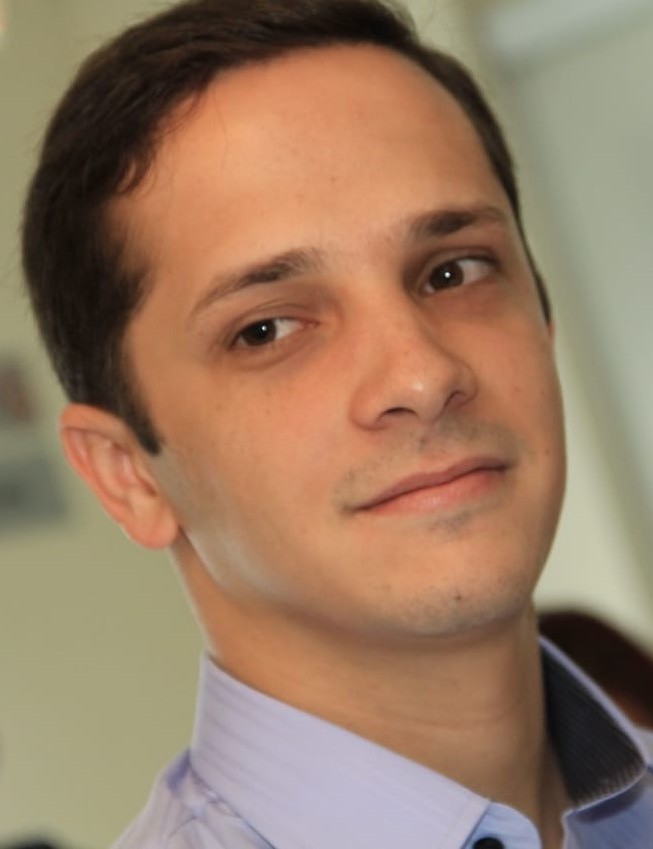}}]{Diego Elias Costa}
  is a postdoctoral researcher in the Department of Computer Science and Software Engineering at Concordia University. He received his PhD in Computer Science from Heidelberg University, Germany. His research interests cover a wide range of software engineering and performance engineering related topics, including mining software repositories, software ecosystems, and performance testing.
  You can find more about him at \url{http://das.encs.concordia.ca/members/diego-costa}.
\end{IEEEbiography}

\begin{IEEEbiography}[{\includegraphics[width=1in,height=1.25in,clip,keepaspectratio]{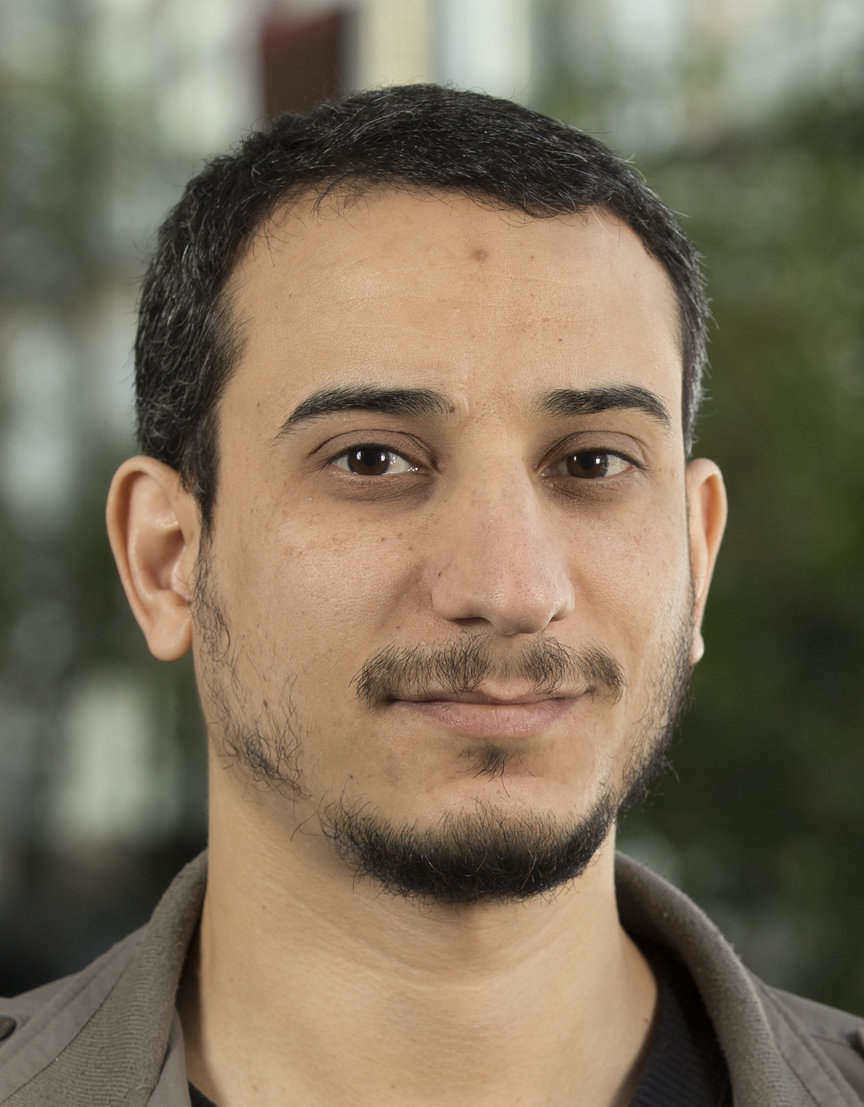}}]{Rabe Abdalkareem}
  is an assistant professor in the School of Computer Science at Carleton University.
  He received his PhD in Computer Science and Software Engineering from Concordia University. His research investigates how the adoption of crowdsourced knowledge affects software development and maintenance. Abdalkareem received his master’s in applied computer science from Concordia University. His work has been published at premier venues such as FSE, ICSME, and MobileSoft, as well as in major journals such as TSE, IEEE Software, EMSE and IST.
  You can find more about him at \url{https://rabeabdalkareem.github.io}.
\end{IEEEbiography}

\begin{IEEEbiography}[{\includegraphics[width=1in,height=1.25in,clip,keepaspectratio]{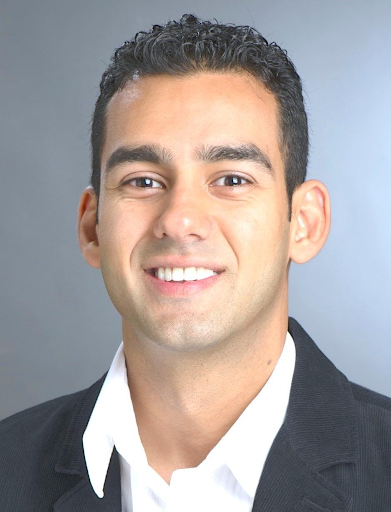}}]{Emad Shihab}
  is an associate professor in the Department of Computer Science and Software Engineering at Concordia University. He received his PhD from Queens University. Dr. Shihab's research interests are in Software Quality Assurance, Mining Software Repositories, Technical Debt, Mobile Applications and Software Architecture. He worked as a software research intern at Research In Motion in Waterloo, Ontario and Microsoft Research in Redmond, Washington. Dr. Shihab is a member of the IEEE and ACM.
  More information can be found at \url{http://das.encs.concordia.ca}.
\end{IEEEbiography}

\begin{IEEEbiography}[{\includegraphics[width=1in,height=1.25in,clip,keepaspectratio]{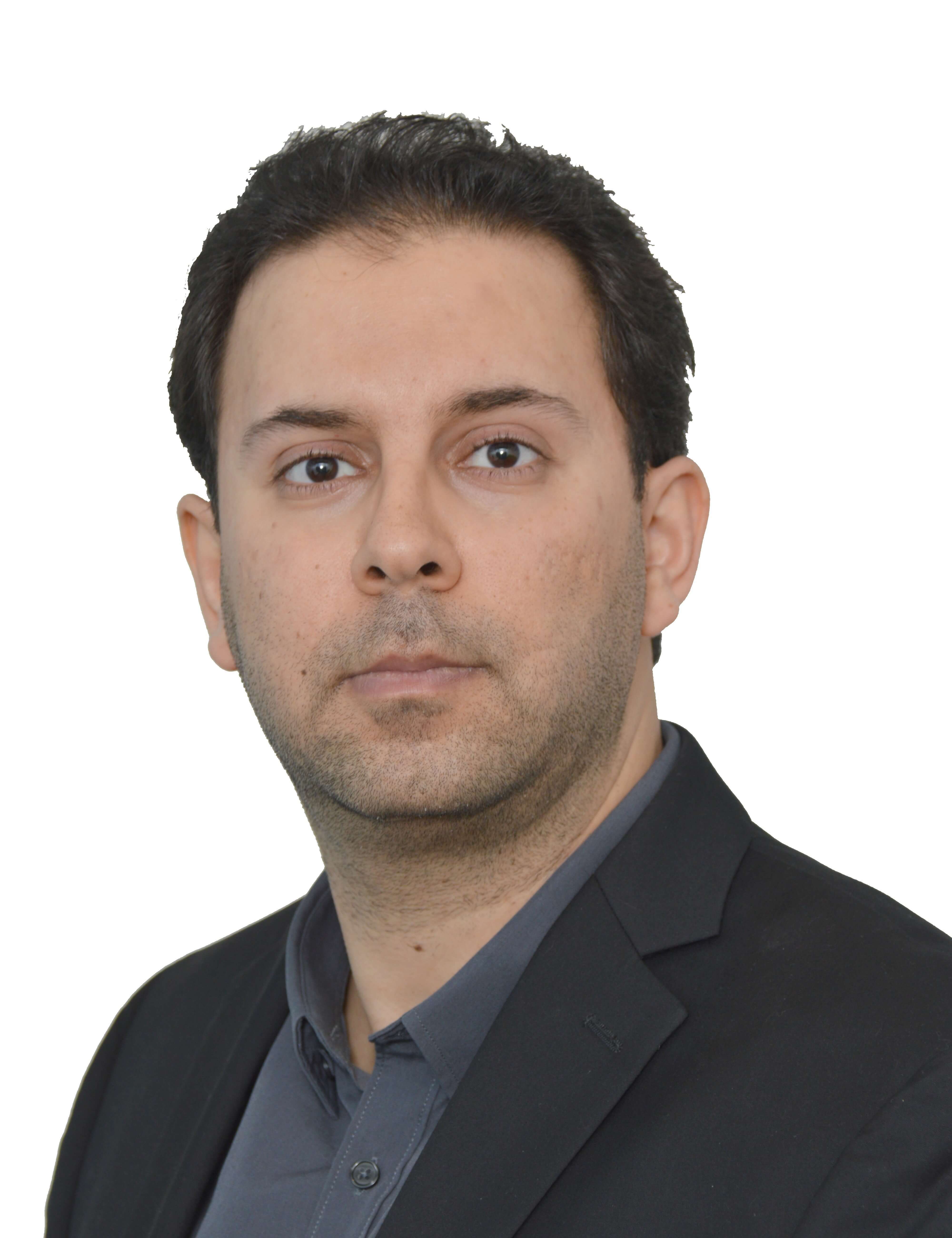}}]{Mohamed~Aymen~Saied}
  is an assistant professor in the department of Computer Science and Software Engineering at University of Laval, Quebec, Canada. Prior to that, he was a Post-Doctoral Fellow in the Electrical and Computer Engineering Department of Concordia University. He received his PhD from University of Montreal. His research interests are at the intersection of software engineering, software data analytics and cloud-native systems.
  You can find more about him at \url{https://saiedmoh.github.io}.
\end{IEEEbiography}

\begin{IEEEbiography}[{\includegraphics[width=1in,height=1.25in,clip,keepaspectratio]{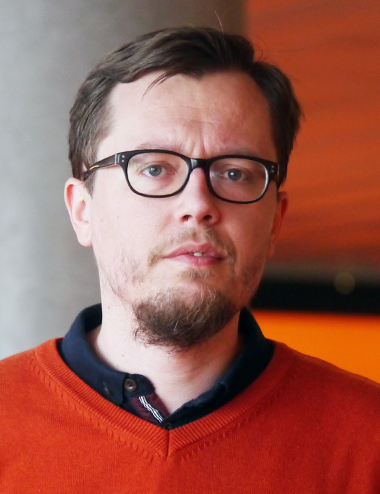}}]{Bram~Adams}
  is an associate professor at Queen's University. He obtained his PhD in 2008 at Ghent University's GH-SEL lab (Belgium). His research interests include software release engineering, mining software repositories, and the role of human affect in software engineering. His work has been published at premier software engineering venues such as EMSE, TSE, ICSE, FSE, MSR and ICSME, and received the 2021 MSR Foundational Contribution Award. In addition to co-organizing the RELENG International Workshop on Release Engineering from 2013 to 2015 (and the 1st/2nd IEEE Software Special Issue on Release Engineering), he co-organized the SEMLA, SoHEAL, PLATE, ACP4IS, MUD and MISS workshops, and the MSR Vision 2020 Summer School. He has been PC co-chair of SCAM~2013, SANER~2015, ICSME~2016 and MSR~2019, and will be ICSE~2023 software analytics area co-chair.
  More information can be found at \url{https://mcis.cs.queensu.ca/bram.html}.
\end{IEEEbiography}

\end{document}